\documentclass[10pt]{iopart}

\usepackage{graphicx}

\begin{document}
\title[Phononic losses in superconducting resonators]{Phononic loss in superconducting  resonators on piezoelectric substrates}
\author{Marco Scigliuzzo$^1$,  Laure E. Bruhat$^1$, Andreas Bengtsson$^1$, Jonathan J. Burnett$^{1,2}$, Anita Fadavi Roudsari$^1$, Per Delsing$^1$}
\address{$^1$Department of Microtechnology and Nanoscience, Chalmers University of Technology, SE-41296, Sweden}
\address{$^2$National Physical Laboratory, Hampton road, Teddington, TW11 0LW, UK}

\ead{scmarco@chalmers.se, per.delsing@chalmers.se}

\begin{abstract}
We numerically and experimentally investigate the phononic loss for superconducting resonators fabricated on a piezoelectric substrate. With the help of finite element method simulations, we calculate the energy loss due to electromechanical conversion into bulk and surface acoustic waves. This sets an upper limit for the resonator internal quality factor $Q_i$. To validate the simulation, we fabricate quarter wavelength coplanar waveguide resonators on GaAs and measure $Q_i$ as function of frequency, power and temperature. We observe a linear increase of $Q_i$ with frequency, as predicted by the simulations for a constant electromechanical coupling. Additionally, $Q_i$ shows a weak power dependence and a negligible temperature dependence around 10$\,$mK, excluding two level systems and non-equilibrium quasiparticles as the main source of losses at that temperature. 
\end{abstract}

\maketitle
\ioptwocol

\section{Introduction}

Superconducting coplanar waveguide (CPW) resonators in the microwave regime have been extensively used in circuit quantum electrodynamics (circuit-QED) because of their high performance (\textit{e.g.} internal quality factors ($Q_i$) can exceed 1 million \cite{megrant2012planar}). Well established and consistent fabrication techniques lead to devices that match its designed parameters, allowing for the implementation of circuits with many multiplexed resonators \cite{Day2003}. Moreover, such resonators have a high spatial confinement of the electric field that enhances the zero point fluctuations, making  the resonators suitable for srtong coupling with other quantum systems \cite{Wallraff2004, goppl2008coplanar}. Nevertheless, energy losses still exist due to energy escaping to the environment. 

Recently, a large effort has been done to understand and reduce the causes of energy loss in superconducting circuits. Primarily, the investigation has focused on geometry \cite{wang2009improving}, substrate and superconductor materials \cite{o2008microwave, sage2011study} as well as fabrication methods \cite{wisbey2010effect, sandberg2012etch}. Additionally, other efforts have explored how wire bonding \cite{wenner2011wirebond} and shielding of the sample box \cite{barends2011minimizing} play a role in the performance of the resonators. Remarkably, one finds the same aforementioned factors responsible for the energy loss of other devices in circuit QED, such as quantum bits (qubits) \cite{dunsworth2017characterization}. This is not surprising since the resonators are fabricated with the same materials, following the same lithographic process, are measured in the same environment and with the same electromagnetic excitation. 

Here, we find that despite well established processing and measurement techniques, CPW resonators fabricated on a piezoelectric substrate are not limited by any of the above factors. This type of substrate has been recently used for building devices that combine circuit QED with mechanical excitations at quantum level \cite{PhysRevX.9.021056, satzinger2018quantum, andersson2019non, chu2018creation}. Interestingly, this technology enables to study semiconductor qubits \cite{ Toida2013,Frey2012} and even to couple them to superconducting qubits \cite{ scarlino2019coherent}. In these hybrid systems, resonators tipically have a $Q_i$ 100 times lower than in comparable devices on a low dielectric loss and non piezoelectric substrate such as silicon or sapphire. Such observations strongly suggest that piezoelectricity introduces an additional loss channel due to its electromechanical coupling. 

In this work, we investigate the mechanical loss channel of superconducting CPW resonators fabricated on gallium arsenide (GaAs). Firstly, we numerically study the dynamics of the mechanical wave generation. Using finite element method (FEM) simulations we show that an oscillating electric field in the resonator leads to the generation of bulk and surface acoustic waves (BAWs and SAWs). A numerical evaluation of the energy loss due to the electromechanical conversion is used to calculate the loss rate and the internal quality factor of the resonator. The simulations predict an unusual linear increase of $Q_i$ with the resonator frequency.  Secondly, we measure the performance of CPW resonators, fabricated on GaAs substrates. We find that the measured quality factors agree well with the numerical simulation, following the predicted linear increase as a function of frequency.


Finally, we observe a weak dependence of $Q_i$ as a function of power and sample temperature. Using a two-level-systems (TLSs) loss model \cite{Burnett_2016} for fitting the response of $Q_i$ as a function of input power, we find that the TLS loss is comparable to those on non-piezoelectric substrates. Furthermore, we measure the temperature dependence of $Q_i$ and the resonance frequency $f_r$. Then, from these fits, we determine the loss rate to quasiparticles (QPs). We find both TLSs and QPs contribute significantly less to losses than the electromechanical conversion.


\section{Method}
\subsection{Design and Fabrication}
\begin{figure}
\centering
\includegraphics[width=0.23\textwidth]{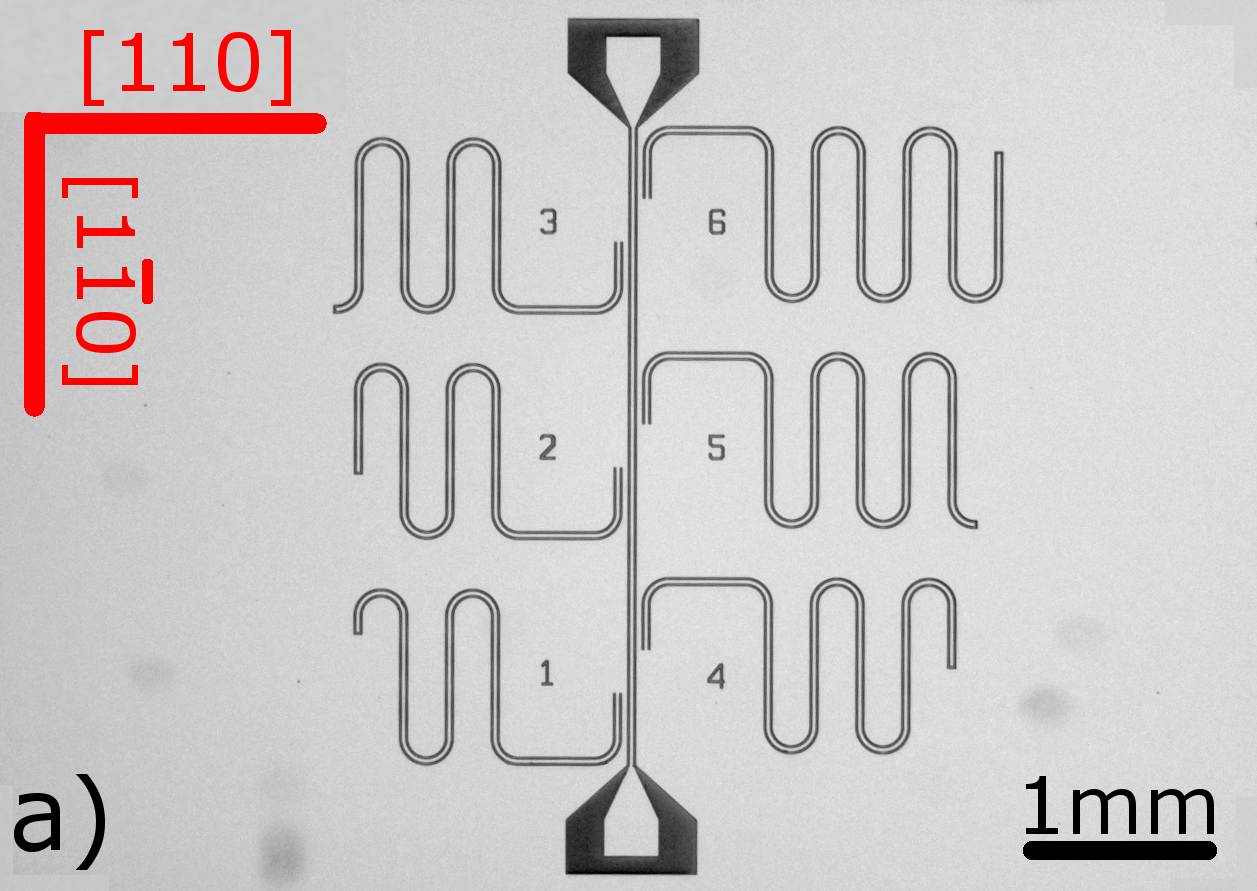}
\includegraphics[width=0.23\textwidth]{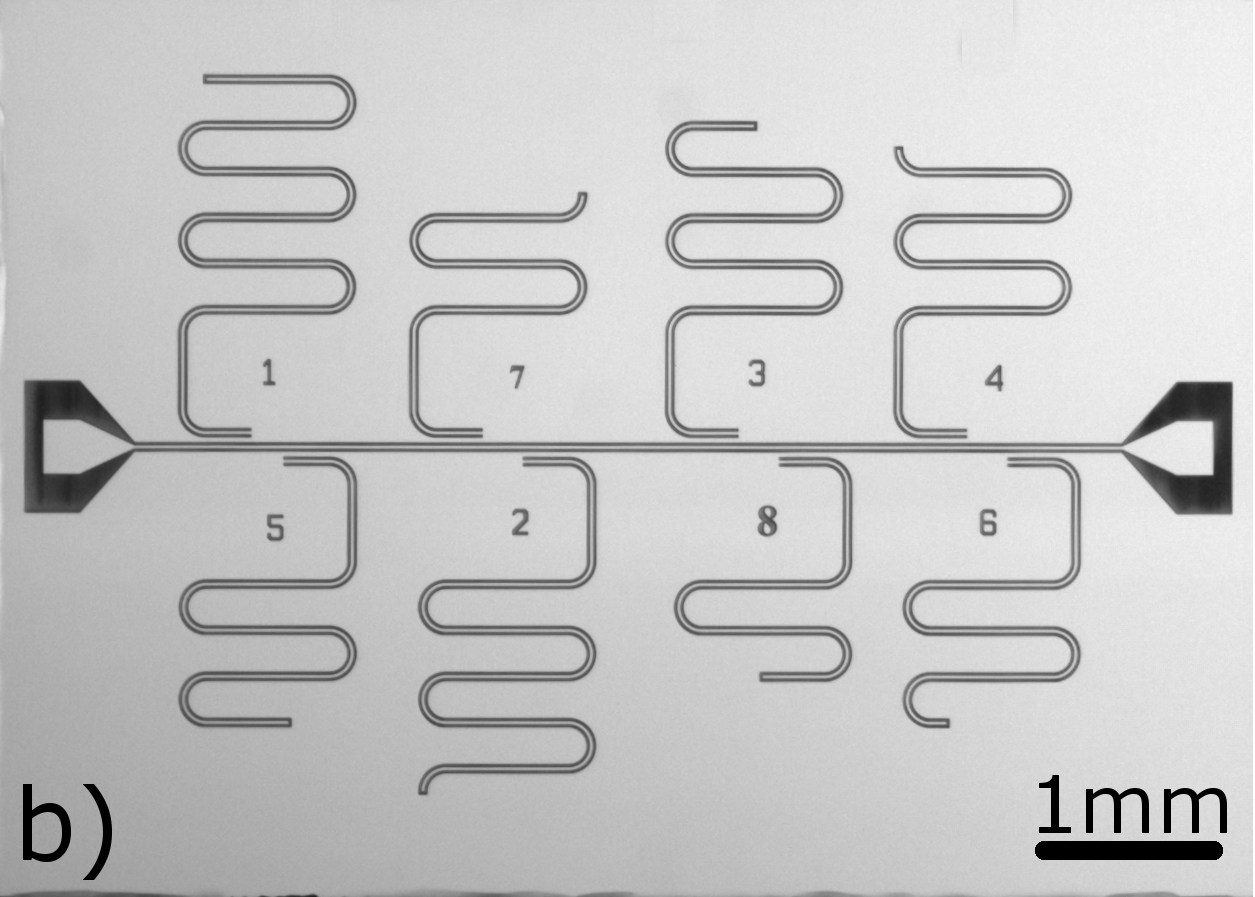}
\includegraphics[width=0.23\textwidth]{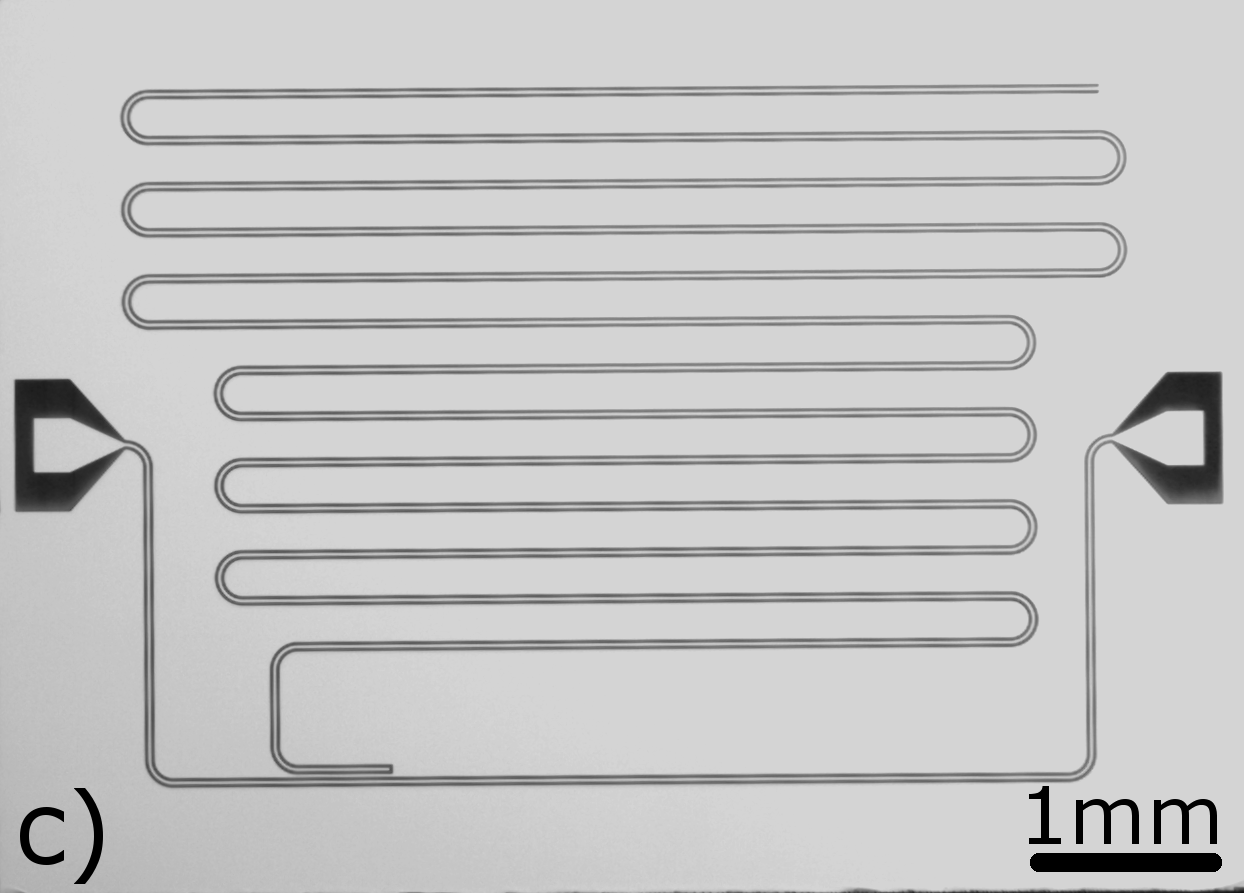}
\includegraphics[width=0.23\textwidth]{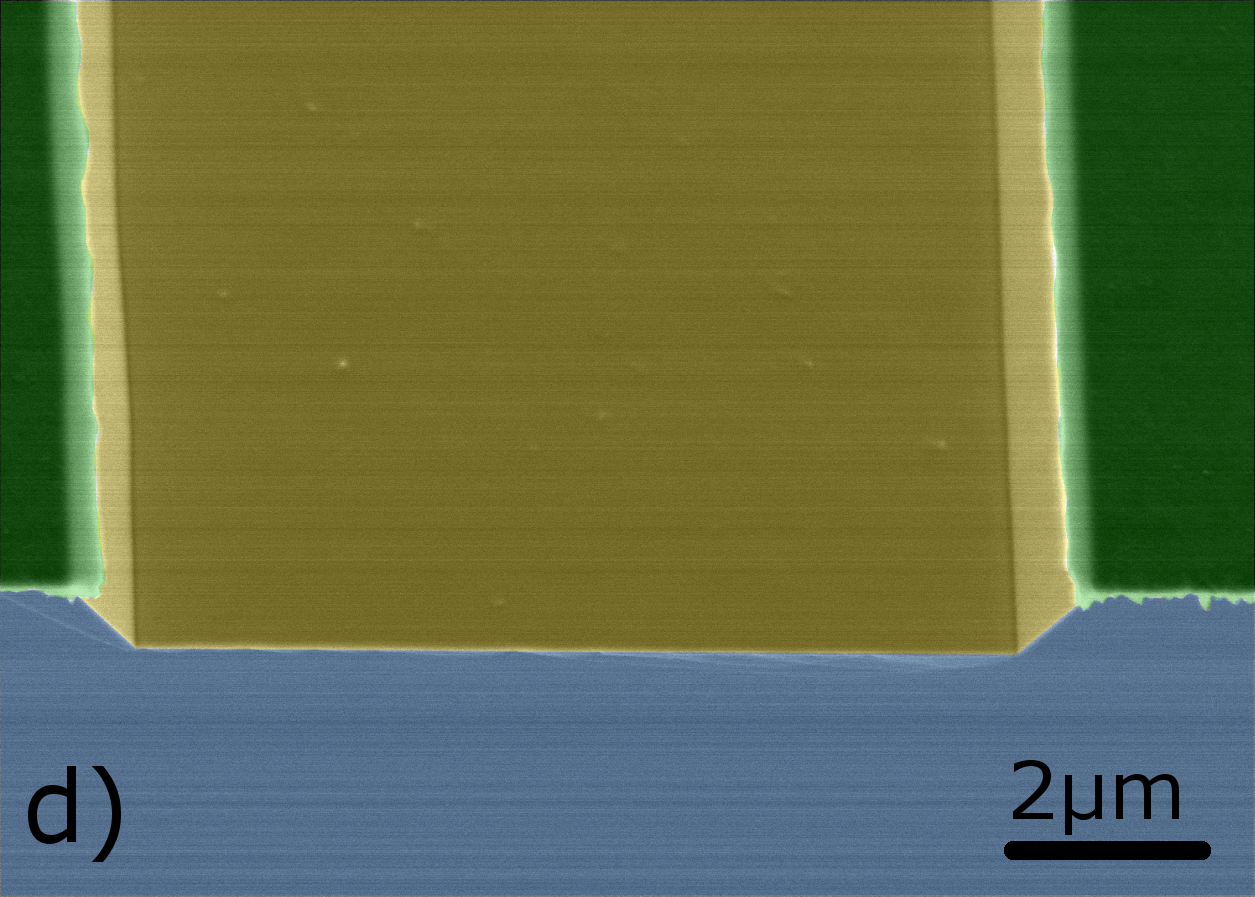}
\caption{The measured devices and a cross-cut of the CPW structure. \textbf{a)} Optical microscope image of a 5x7\,mm$^2$ chip of the type Vertical: the brighter part is the 150\,nm aluminium film deposited on the darker GaAs substrate. The crystallographic direction of the GaAs lattice are highlighted in red. \textbf{b)} Optical microscope image of a Horizontal chip. \textbf{c)} Optical microscope image of a Harmonics chip. \textbf{d)} False color scanning electron microscope image of the gap (GaAs, yellow) between the ground plane and the center conductor (aluminium, both green). In the gap area, the GaAs substrate is etched to a depth of about 1-2$\,\mu$m, creating trenches that slightly increase the resonator performance.}
\label{fig:fab}
\end{figure}
The measurements are performed on three different designs of quarter wavelength ($\frac{\lambda}{4}$) CPW resonators. The first geometry (figure \ref{fig:fab}a), labelled Vertical 1 and Vertical 2, consists of two nominally identical chips, each contains six resonators with different frequencies. The second geometry (figure \ref{fig:fab}b), labelled Horizontal 1 and Horizontal 2, also consists of two nominally identical chip with eight different resonators. However, these chips have a perpendicular orientation with respect to the Vertical samples (the crystallography directions are shown in the figure \ref{fig:fab}). All resonators are multiplexed in a notch geometry with inductive coupling to a feedline. Their length is chosen such that the fundamental mode of each resonator lies in the 4-8\,GHz band. The last design, sample Harmonics, consists of a single long multimode-resonator, with the fundamental resonance frequency  at 450\,MHz, and a capacitive coupling to the feedline. The main features of all devices are summarized in table \ref{tab:devices}.

The resonators are fabricated with a common lithography process on the (100) polished surface of a commercial epi-ready 2-inch GaAs wafer. The wafers come individually packaged in an air tight plastic package. Within the cleanroom, the package is opened and immediately placed in the load lock chamber of an electron beam evaporator (Plassys MEB550S) and pumped to a pressure of $10^{-7}\,\mbox{mbar}$. While pumping the wafer is heated to 300\,$^o\mbox{C}$ and kept at this temperature for $10\,\mbox{min}$. The wafer cools down while the system reaches its base pressure of  $4\cdot10^{-8}\,\mbox{mbar}$ which takes 10 hours. A layer of 150\,nm of 99.999\% (5N) pure aluminium is evaporated while the wafer rotates around its axis to increase the film homogeneity. The aluminium surface is oxidized after the evaporation by 10\,mbar static pressure of 99.99\% (4N) pure oxygen in the chamber. After removing the metalized wafer from the Plassys MEB550S, a 1.3\,$\mu$m thin AZ1512 resist layer is spun on the wafer and baked at 112\,$^o\mbox{C}$. The resist is then patterned with a laser writer (Heidelberg DWL2000). Prior to development, an additional thick protective layer of the same resist is spun on the back side of the wafer and baked in an oven at 110\,$^o\mbox{C}$ for 2\,min. The extra resist prevents the production of gallium arsenide dust due to etching of the back side of the wafer. The patterned resist is then developed with AZ dev:H$_2$O 1:1 solution followed by a mild ashing in an oxygen plasma. The design is then wet etched with the commercially available Transene type A aluminium etchant. When the exposed aluminium is etched away, the unprotected GaAs surface beneath is rapidly etched by the acid solution, as shown in figure \ref{fig:fab}d,  creating trenches of 1-2\,$\mu$m under the gaps in the resonator electrodes. 

 
\begin{table}[]
    \centering
     \caption{The five measured devices and their most important features are summarized. The corresponding chips are shown in figure \ref{fig:fab}. All chips are made of aluminium on gallium arsenide. The orientation refers to the direction of the longest straight segment in the meandering part of the resonators. In the last column the main coupling mechanism to the feedline is reported.}
    \label{tab:devices}
    \tabcolsep=0.13cm
    \small
  \begin{tabular}{c c c c c}
  \hline
 \vspace{0.1cm}
 Sample & figure &  orientation &  trench $(\mu m)$ & coupling \\ 
 \hline 
 \vspace{0.1cm}
  Vertical 1 & \ref{fig:fab}a & [110] & $1.5\pm0.3$ & inductive \\  [0.1ex] 
  \vspace{0.1cm}
 Vertical 2 & \ref{fig:fab}a & [110] & $1.5\pm0.3$ &inductive \\[0.1ex] 
  \vspace{0.1cm}
 Horizontal 1 & \ref{fig:fab}b & $[1\overline 1 0]$ & $0.6\pm0.2$ &inductive \\[0.1ex] 
 \vspace{0.1cm}
 Horizontal 2 & \ref{fig:fab}b & $[1\overline 1 0]$ &$0.8\pm0.2$ &inductive \\[0.1ex] 
 \vspace{0.1cm}
 Harmonics & \ref{fig:fab}c & $[1\overline 1 0]$ & $0.8\pm0.2$ &capacitive \\ [0.1ex] 
\end{tabular}
   
\end{table}

\subsection{Experimental Setup}
\begin{figure}
\centering
\includegraphics[width=0.34\textwidth]{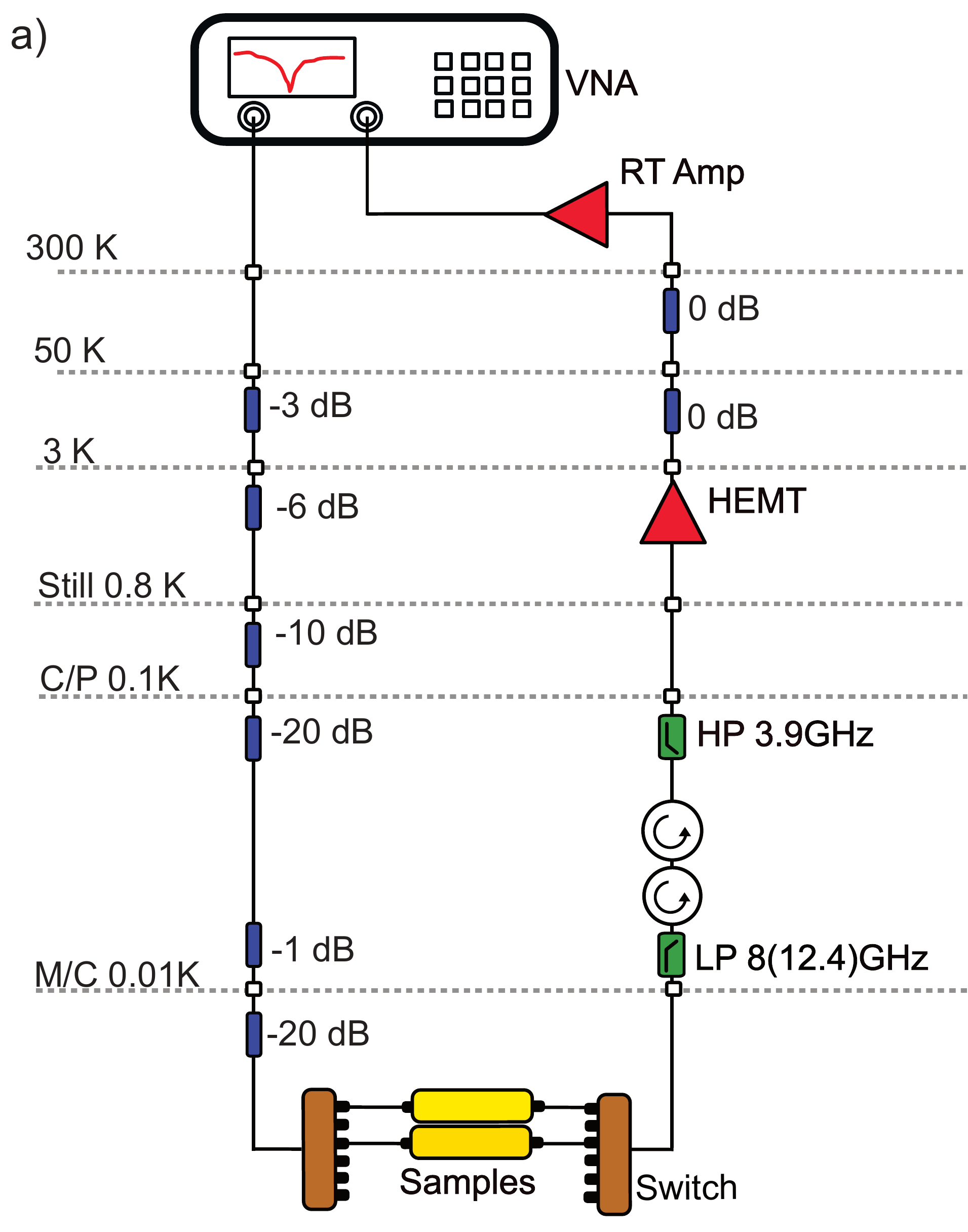}
\includegraphics[width=0.23\textwidth]{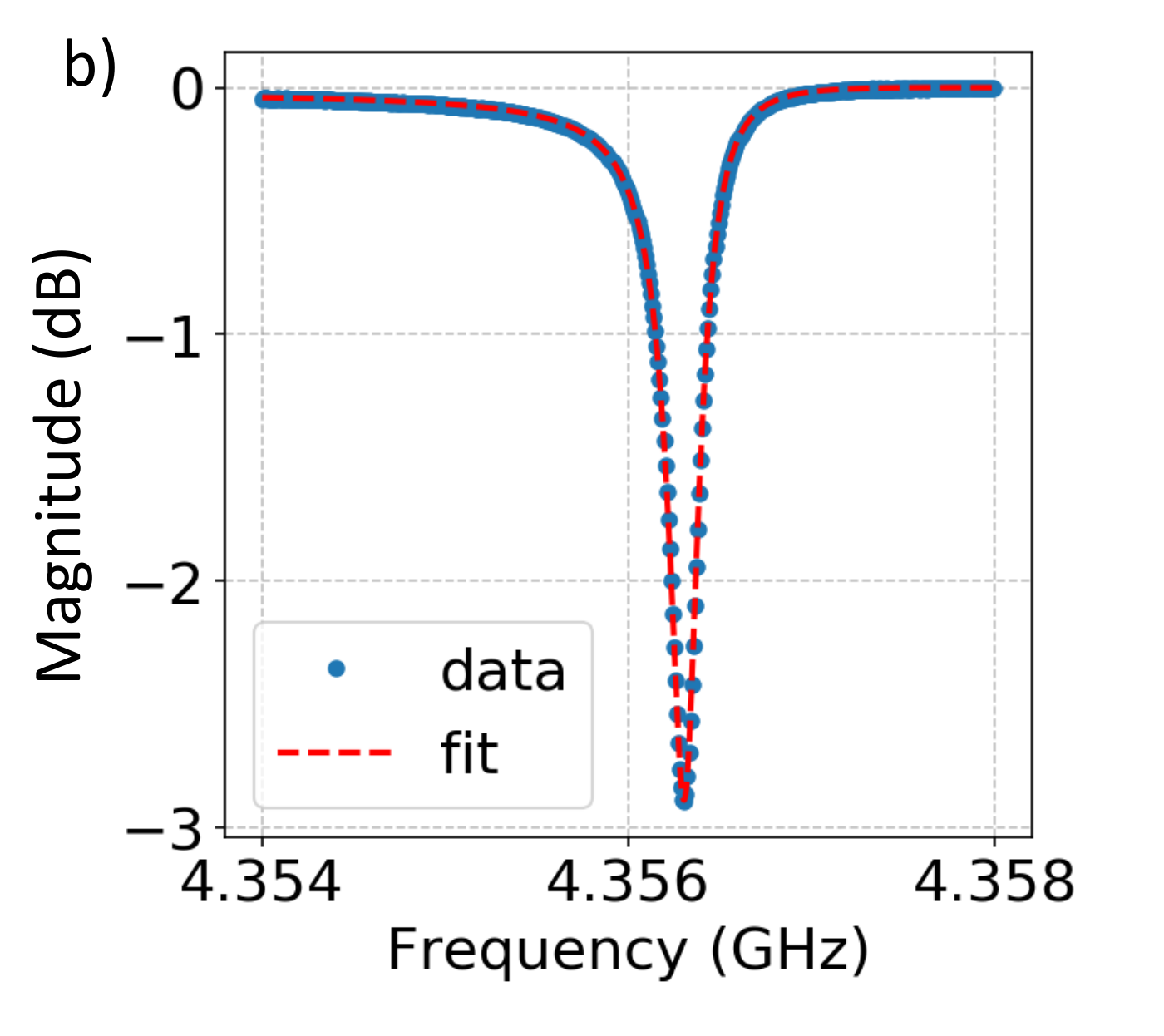}
\includegraphics[width=0.24\textwidth]{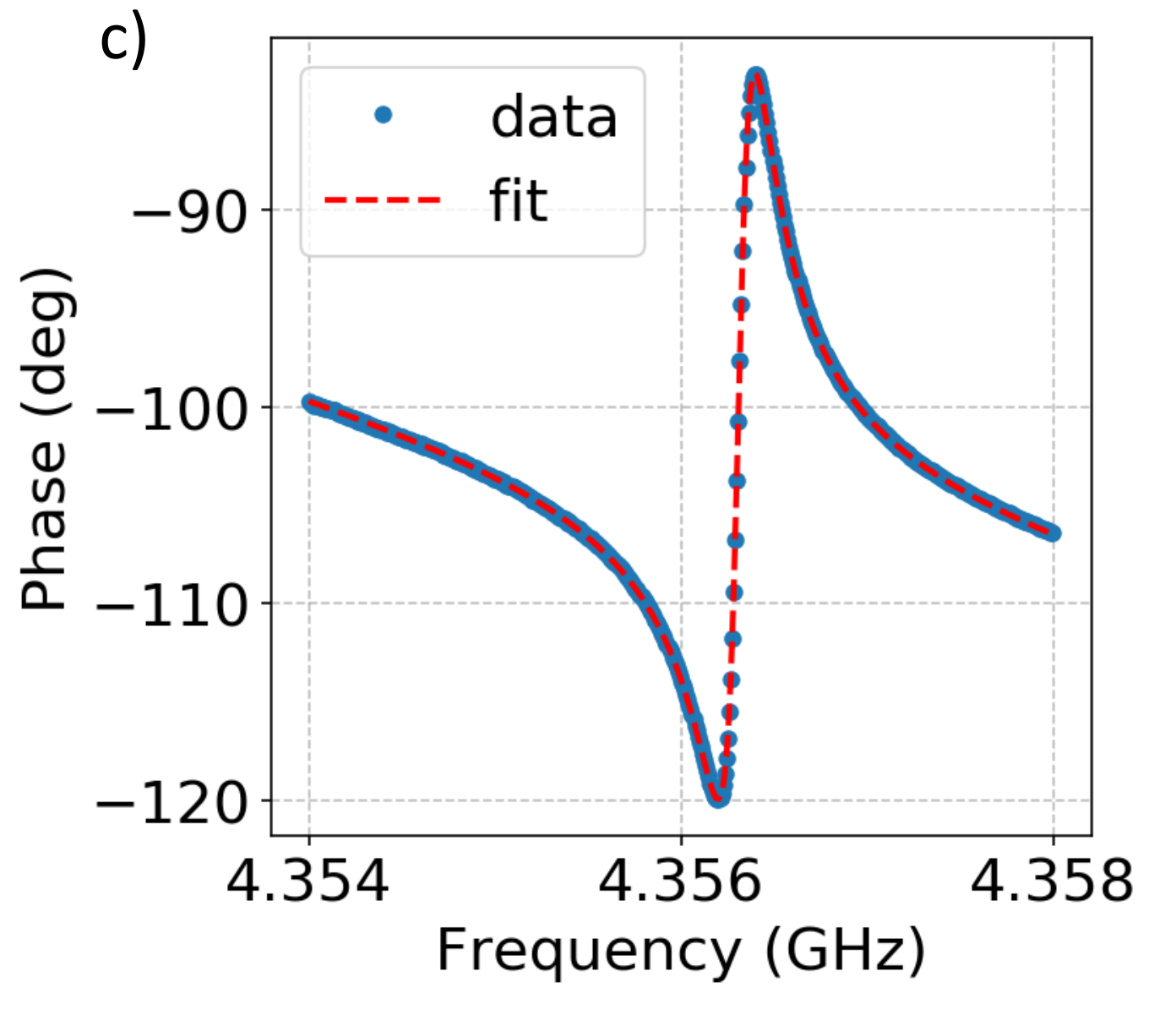}
\caption{\textbf{a)} Cryogenic and room temperature setup used in the resonator measurement. The presence of microwave switches at the mixing chamber stage allows the measurement of multiple chips during the same cooldown. \textbf{b)}  Magnitude and \textbf{c)} phase as function of frequency of the transmitted signal measured for one of the resonators of the sample Vertical 1 (blue dots). $Q_i$ is extracted with a global fit in quadrature space (red dashed line).}
\label{fig:setup_fit}
\end{figure}

The chips are wire-bonded in an oxygen-free copper sample-box with aluminium wire. The sample-box is thermally and mechanically anchored at the mixing chamber stage of a Bluefors LD250 dilution refrigerator and cooled to a base temperature of 10\,mK (see figure \ref{fig:setup_fit}a). 
Two different setups are used to measure the resonators. For samples Vertical 1 and Vertical 2 a narrow-band 4-8\,GHz HEMT amplifier is used. Samples "horizontal 1", "horizontal 2" and "harmonics" are measured with a wider band 3-12\,GHz HEMT amplifier and wider circulators. After an additional stage of room temperature (RT) amplification, their transmission spectra are measured with a Vector Network Analyzer (VNA).  

The resonators $Q_i$, together with the coupling quality factor $Q_c$ and the resonance frequency, are extracted using the circle fit routine \cite{probst2015efficient}. The measured and fitted amplitude and phase of one resonator transmission spectrum (sample Vertical 1) are shown in figure \ref{fig:setup_fit}. The routine globally fits the imaginary and real part of the resonator spectrum.


\subsection{Numerical Modelling}
An experimental study of the mechanical waves generated by the resonators is difficult. 
Firstly, even a relatively large level of excitation, corresponding to $10^{9}$ photons in the resonators or an electrical energy $E_{int}\approx 10^{-15}$\,J (see Eq. \ref{for:single}), produces a small potential between the superconductor electrodes $\approx d\sqrt{\frac{E_{int}}{V}}$ of the order of $10^{-5}$V, where $V$ is the volume that confines the electric field and $d$ is the distance between the resonator electrodes. This potential generates displacements that do not exceed $s=d\max\left[\frac{e_{ij}}{ c_{ij}}\right]E\approx10^{-17}$\,m, where $e_{ij}$ is the piezoelectric tensor, $c_{ij}$ the elasticity tensor and $E$ is the electric field. Although there are techniques capable of detecting such a small displacement \cite{knuuttila2005laser, gustafsson2012local}, they are very hard to use at 10\,mK, over a large area and within a closed sample-box. 


We would like to remark that the largest average photon number that we are able to excite the cavity is slightly below $10^9$ (see figure \ref{fig:pow_dependance}). Excitation levels with much lower photon number generate waves that cannot be detected with the mentioned methods. 

Instead, in order to study and predict the behaviour of the resonators we develop a numerical model of the system. The model is realized using a commercial finite element analysis software  (COMSOL Multiphysics, version 5.3a, Piezoelectric module, time dependent study). We use this model to calculate the strain tensor $s_{ij}$ and the velocity $v$ at each point in a grid that discretizes the substrate, given an electrical excitation as initial condition. From these quantities it is possible to calculate the mechanical energy released into the substrate:  
 \begin{equation}
     E_{mech}=\int_{V}dV\frac{1}{2}\rho  v^2+\frac{1}{2}c_{ij}s_{ij}^2,
 \end{equation}
 where $\rho$ is the substrate density and $V$ its volume.
 
This calculation can be repeated at each time during the evolution of the system, and its time derivative represents the mechanical energy loss rate $L_{\mbox{\small{mech}}}$. As already mentioned, other sources of energy loss may contribute to the $Q_i$ of the resonator, in particular TLS and QP loss rates, denoted $L_{\mbox{\small{TLS}}}$ and $L_{\mbox{\small{QP}}}$ respectively. The internal $Q_i$ value is defined and divided into different terms as: 
\begin{equation}
\centering
\eqalign{
\frac{1}{Q_i} &=\frac{1}{2\pi}\frac{\mbox{energy lost per cycle}}{\mbox{energy stored}} \\
        &=\frac{1}{2\pi f}\frac{L_{\mbox{\small{mech}}}+L_{\mbox{\small{TLS}}}+L_{\mbox{\small{QP}}}+L_{\mbox{\small{residual}}}}{\mbox{energy stored}} \\
        &= \frac{1}{Q_{mech}}+ \frac{1}{Q_{TLS}}+ \frac{1}{Q_{QP}}+ \frac{1}{Q_{\mbox{\small{residual}}}},
}
\label{eq:Q_internal}
\end{equation}
where additional unknown sources of losses have been taken into account in a residual loss rate $L_{\mbox{\small{residual}}}$, and the corresponding quality factor $Q_{\mbox{\small{residual}}}$. We need to take all of them into account so that we can determine the dominant one. 

A full scale 3-dimensional (3D) model is intractable on a desktop computer: the full chip is 5x7x0.35\,mm$^3$, and the mechanical wavelength at 6\,GHz is about 0.5\,$\mu$m. A good discretization would require at least $10^{14}$ elements, each with the mechanical and the electric degrees of freedom. In order to use a desktop computer we can instead exploit the cubic symmetry of the GaAs elastic tensor, and the tranlational invariance of the CPW resonator. The mechanical displacement produced by the transverse electric magnetic (TEM) mode in the resonator is confined in the transverse plane when the resonators electrodes direction is [110] or [$1\overline10$]. Thus we can reduce our model to a two dimentional (2D) simulation \cite{doi:10.1177/1045389X18803461}.

Our model geometry consists of a 2D section perpendicular to the CPW resonator and the  substrate surface near the electrode (the blue surface in the false color SEM picture in figure \ref{fig:fab}d). The metallic layer of the electrodes and the vacuum are also included in the simulation.

The maximum element dimension for the mesh is dynamically adapted as a function of the simulation frequency: the wavelength $\lambda$ of SAWs and BAWs can be calculated from their speed in GaAs for each frequency, and the element mesh size, $\delta x$, does not exceed 0.1\,$\lambda$.  

The electric potential on the metallic layer is changed periodically following a harmonic oscillation $V=V_0\sin(2\pi f t)$. The simulation time consists of 10 oscillation periods, and the time discretization $\delta t$ follows the Courant condition \cite{1967IBMJ...11..215C}, \textit{i.e.} the numerical speed $\frac{\delta x}{\delta t}$ is larger than the wave propagation velocity. During this time the mechanical waves generated do not reach the edge of the substrate because the total dimension of the substrate is dynamically adapted to the frequency of the simulation to prevent this possibility.

\subsection{Simulation Limitations}
The five orders of magnitude difference between the phononic and photonic wavelength limits the dimensions of the device (or device portion) that can be simulated. As already mentioned, we cannot simulate the full chip, and not even a single resonator in 3D. In a case where the elastic and the piezoelectric tensor have less symmetry than GaAs, our approach does not work and instead quasi-3D simulations are needed. 

The large difference between light and sound speed allows us to simplify the full electromagnetic problem into an electrostatic one. The electrostatic equations are coupled to the mechanical dynamics ones, such that a solution is found for the electrostatic part and updated instantly to the full domain. This is justified since the slow mechanical dynamics behaves as quasi-static compared to the time scale of electromagnetic evolution.

Finally our simulations show that the steady state regime in the energy conversion is reached after a few oscillations. Nevertheless, this is not the real system steady state, because we are ignoring the back scattering at the boundaries of the chip. Unfortunately we cannot take this into account due to the dimensions of the substrate. 

\section{Simulations Results}
\subsection{Bulk and Surface Acoustic waves conversion rate}
\begin{figure}
\centering
\includegraphics[width=0.44\textwidth]{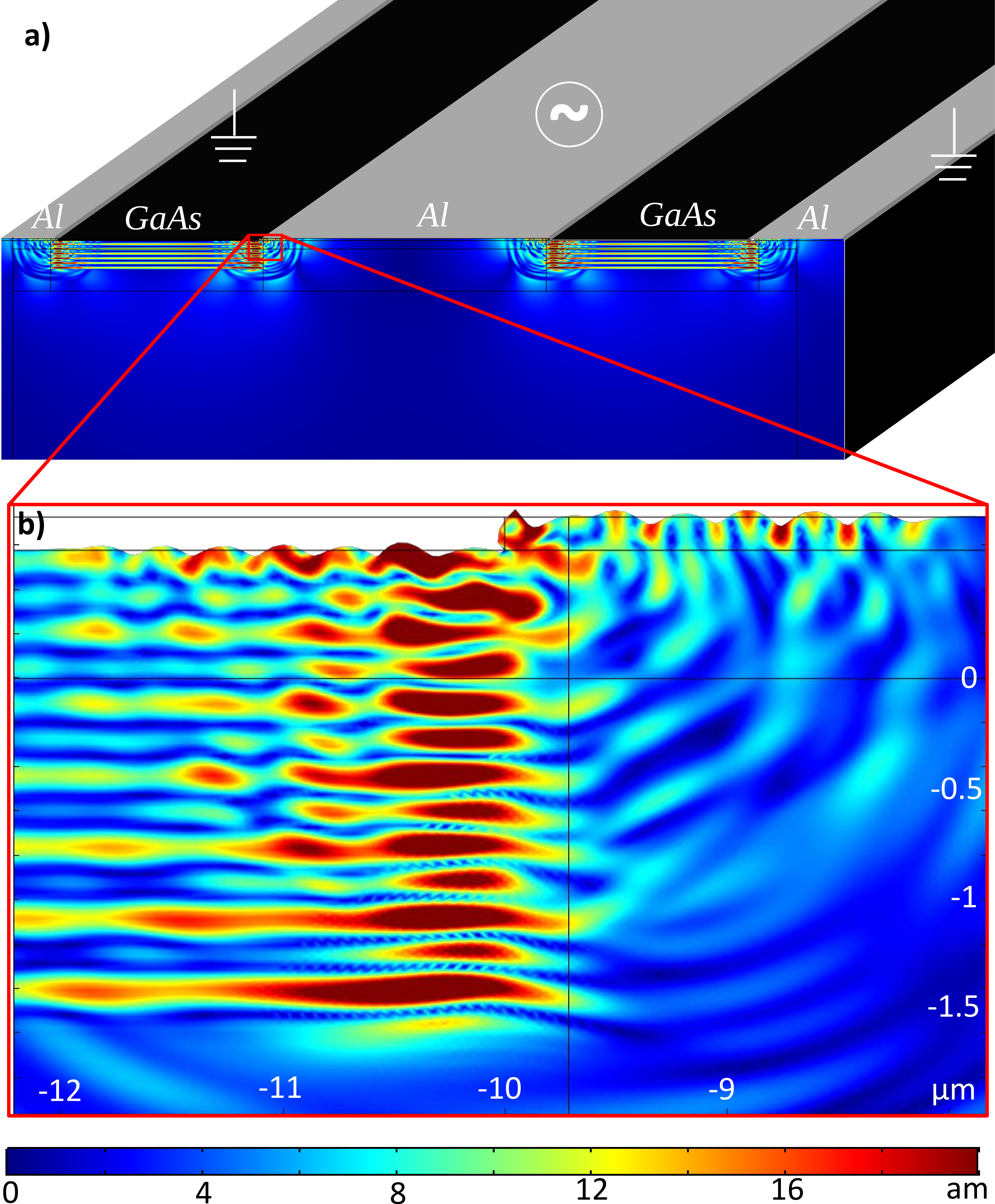}
\caption{\textbf{a)} Displacement of the substrate and the metal layer six periods after the 10\,GHz electrical oscillation is applied. Following the colour scale, the points of larger displacement from the equilibrium condition are located in the red area, while the blue area represents the points that are at rest position. \textbf{b)} Zoom in of the edge of the electrode (red rectangle in a)) highlights the piezoelectric conversion of electric field into BAWs generated on the surface propagating inside the substrate, while SAWs travel on the surface. The deformation in the plot has been amplified by $2\cdot10^9$ times to make it visible.}
\label{fig:simulation_displacement}
\end{figure}
The main results of the simulation is the calculation of the electrical energy converted into mechanical strain energy and kinetic energy. The total mechanical energy is subtracted from the electromagnetic energy stored in the resonator and it leaves the area close to the electrodes as surface and bulk acoustic waves: this is the mechanical part of the loss rate of the resonator. 

Figure \ref{fig:simulation_displacement} shows the displacement of the substrate from its rest position six periods after the electric oscillating potential was applied to the electrodes. The larger displacements are due to BAWs that travel inside the substrate. The SAWs generated travel along the surface and the larger deformation is located in and under the metallic layer. 

\begin{figure}
\centering
\includegraphics[width=0.47\textwidth]{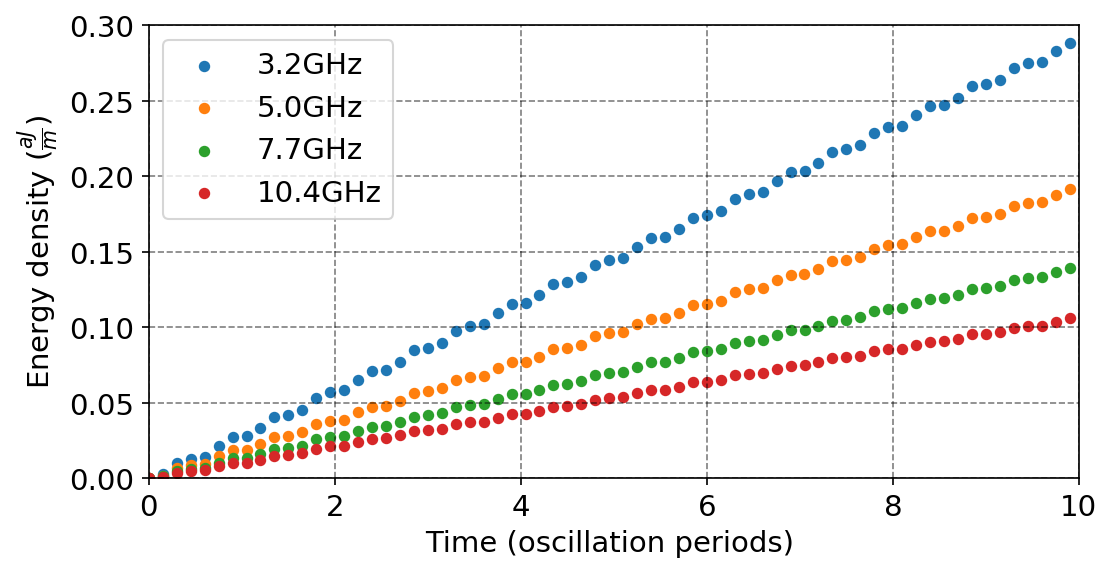}
\caption{Numerical evaluation of the total mechanical energy density released in the substrate as acoustic waves (SAWs and BAWs), versus the number of oscillations of the electric field at different frequencies.  The calculation of $Q_i$ takes into account the steady state slope of the electromechanical conversion rates, and it is extracted with a linear fit. The SAWs represents 2\% of the total mechanical energy released into the substrate.}
\label{fig:loss_rate}
\end{figure}

 By integrating the mechanical energy density on the substrate domain it is possible to estimate the loss rate due to the piezoelectric effect (see figure \ref{fig:loss_rate}). The dominant loss is due to BAWs, the energy converted to SAWs represents only 2\% of the total mechanical energy, and can be estimated by integrating only in a two wavelengths thick superficial layer, excluding the zone right below the CPW gap. 

It is interesting to notice that the energy lost in an oscillation period decreases with frequency. This is easily understood if we consider that the electromechanical coupling is almost constant in the GHz range: on one hand, the energy converted in mechanical excitation has a constant rate but as the frequency increases the period gets shorter leading to a reduction of total energy lost per period. On the other hand, the electric energy stored in the resonator does not change with frequency. Therefore the $Q_i$, their ratio, increases as a function of frequency. 

The expected linear increase of the internal quality factor of a quarter-wavelength CPW resonator as a function of frequency \cite{Pozar2011} is usually overcome by a corresponding increase in the intrinsic loss tangent. 

We simulate the wave generation with and without trenches under the electrodes gap, and calculate the conversion rates for different frequencies. The derivative of these loss rate is used to estimate the $Q_{mech}$, and the results are shown in figure  \ref{fig:Qi}. From this quantity, we can extract the $Q_i$ as shown in equation \ref{eq:Q_internal}. We find an almost linear increase of the internal Q-factor with respect to frequency.

\section{Experimental Results}
\subsection{Frequency dependence of $Q_i$}
\begin{figure}
\centering
\includegraphics[width=0.47\textwidth]{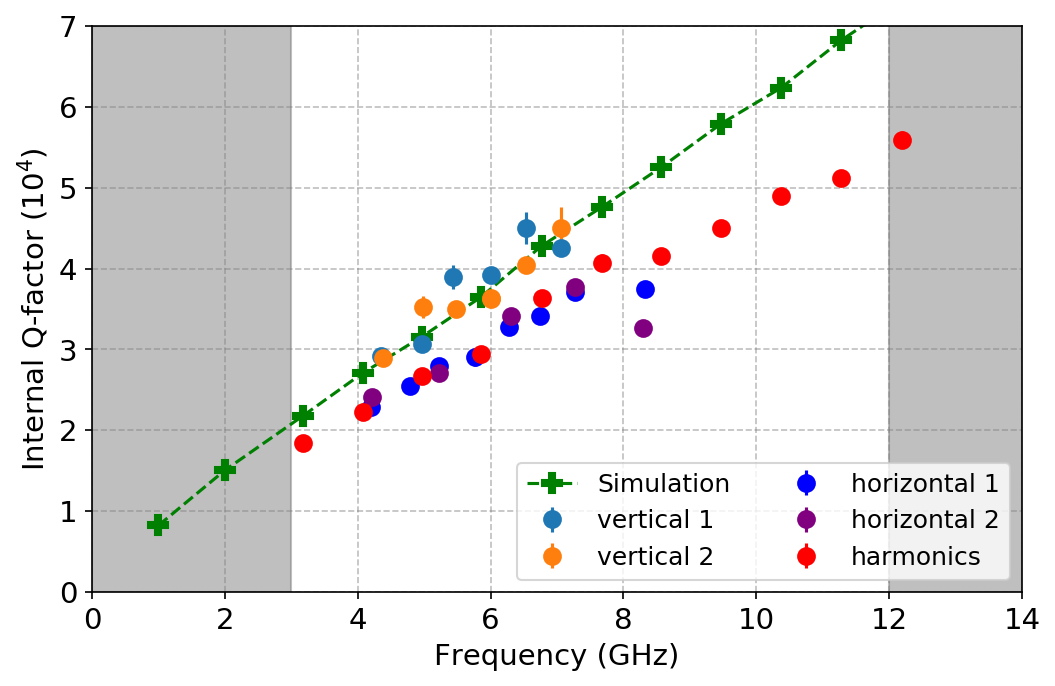}
\caption{Internal Q-factor (at $10^7$ photons stored in the resonator) of the resonators fabricated on different samples and the values extracted from the simulations. Each point of sample Vertical 1 and 2, Horizontal 1 and 2, represents Q-factor at the fundamental resonance of each resonator, while the data of sample Harmonics are the internal Q-factor of higher modes of a low frequency resonator. The green markers show the numerical results for the Q-factor for resonators with trenches $1\,\mu$m deep. The grey regions represent the band outside the HEMT amplifier bandwidth.}
\label{fig:Qi}
\end{figure}

The main results are shown in figure \ref{fig:Qi}, where the internal $Q_i$ is measured for the 5 different samples and compared with the simulations results. Samples Vertical 1 and 2 are fabricated with 90 degrees orientation compared to samples Horizontal 1 and 2. In figure \ref{fig:Qi} we plot the internal Q of their fundamental mode. Sample Harmonics is a long multimode resonator, and in the plot we show the $Q_i$ of its different harmonics. 


The predicted increase of $Q_i$ as a function of frequency is measured for all samples; the not perfect homogeneity among different resonators on a chip produces the scatter across the samples. In fact, nominally identical samples show a variation of similar magnitude. However, sample Harmonics is a single multimode resonator, so all the values $Q_i$ for different frequencies are measured on the same resonator. Moreover, with this design, we have a free spectral range small enough to sample the full bandwidth of our measurement setup. The $Q_i$s of the different modes show again a good agreement with the simulation results. 

We expect no difference between the three geometries because of the symmetry of the GaAs lattice. The electric field produces the same mechanical displacement for the two orientations. Nevertheless in figure \ref{fig:Qi} we can see that the Vertical samples present a sligthly larger internal Q factor. We attribute the variation of the $Q_i$ to deeper trenches on the samples Vertical. 

\subsection{Power dependence of $Q_i$}
\begin{figure}
\centering
\includegraphics[width=0.47\textwidth]{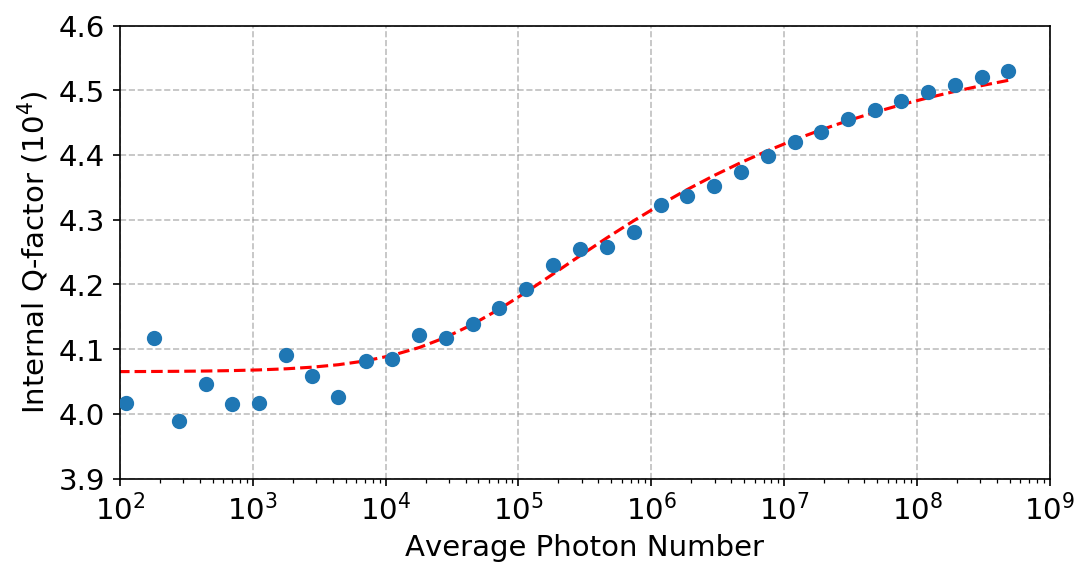}
\caption{Internal Q-factor as a function of average photon number occupation for the resonator with fundamental mode $f=6$\,GHz on sample Vertical 1. The red dashed line shows a  fit to a TLS power dependence. The weak change in Q indicates that the TLS have only a small contribution to the total loss rate even at the lowest powers.}
\label{fig:pow_dependance}
\end{figure}

Additional information on the nature of the loss can be inferred by studying the response of the resonator as a function of driving power ($P$) . In figure \ref{fig:pow_dependance}  we plot the internal quality factor with respect to the average photon number $ \left\langle n\right\rangle$ in the resonator. Its value is calculates as:
\begin{equation}
    \left\langle n\right\rangle=\frac{\left\langle E_{int}\right\rangle}{hf_r}=\frac{1}{\pi}\frac{Z_0}{Z_r}\frac{Q_l^2}{Q_c}\frac{P}{hf_r^2},
    \label{for:single}
\end{equation}
where $\left\langle E_{int}\right\rangle$ is the average electromagnetic energy stored in the resonator, $f_r$ is the resonance frequency, $Z_0$ ans $Z_r$ are the environment impedance (the feed transmission line) and the resonator impedance  respectively. Both are designed to be close to $50\Omega$ (the trenches will decrease the value of capacitance per unit length, resulting in a slightly larger impedance, for both resonator and transmission line). $Q_l$ and $Q_c$ are the loaded and the coupling Q of the resonator.

The common TLS loss model \cite{Burnett_2016} states that at high power (average number of photons much larger than the critical photon number, $\left\langle n\right\rangle\gg n_c$) the TLS loss rate becomes negligible compared to other losses. Moreover, a large temperature ($T$) compared with the resonator frequency can saturate the TLSs reducing their loss rate. The TLS loss as function of $\left\langle n\right\rangle$ and $T$ can be modelled as:
\begin{equation}
    \frac{1}{Q_{i}}=\frac{1}{Q_{TLS}^0}\frac{\tanh\left(\frac{hf_r}{2k_BT}\right)}{\left(1+\frac{\left\langle n\right\rangle}{n_c}\right)^\beta}+\frac{1}{Q_{A}},
    \label{eq:TLS_model}
\end{equation}
where $Q_{TLS}^0
$ is the Q-factor due to the TLS at very low temperature and photon number, and $Q_{A}$ represents the Q due to all losses other than TLSs. The energy loss is due mainly to resonant TLSs, while non-resonant ones contribute to a frequency shift of the resonator frequency with respect of temperature \cite{pappas2011two}).

From the fit shown in figure \ref{fig:pow_dependance}, we obtain  $Q_{TLS}^0=3.5\pm0.1\cdot10^5$, close to the literature value for CPW resonators on low loss non-piezoelectric substrates such as silicon or sapphire. However, the value extracted for the high photon number is $Q_{A}=4.6\pm0.5\cdot10^4$, very far from the common values ($\approx10^6$). This value is instead very close to our simulation of internal $Q_i$ from acoustic losses. The value $\beta=0.2\pm0.02$ is similar to those found in literature for non-piezoelectric substrates\cite{Burnett_2018, PhysRevApplied.11.044014}. A summary of these values for the different samples is reported in table \ref{tab:TLS}.

\begin{table}[]
    \centering
    \caption{Average values and maximum dispersion errors resulting from the best fit function with formula \ref{eq:TLS_model}. $Q_{TLS}^0$ is the internal quality factor due to the TLSs loss, $n_c$ is the average critical photon number and $\beta$ is the exponent as shown in Eq. \ref{eq:TLS_model}.} 
  \begin{tabular}{c c c c} 
\hline
 \vspace{0.1cm}
 Sample & $Q_{TLS}^0 (10^{5})$ & $\beta$ & $n_c (10^3)$ \\ 
 \hline 
 \vspace{0.1cm}
  Vertical 1 & $3.0\pm1.4$ & $0.19\pm 0.03$ & $6\pm 6$ \\  [0.1ex] 
  \vspace{0.1cm}
 Vertical 2 & $3.3\pm1.6$ & $0.19\pm 0.06$ &$10\pm8$ \\[0.1ex] 
  \vspace{0.1cm}
 Horizontal 1 & $1.2\pm0.5$ & $0.21\pm0.02$ & $4\pm2$ \\[0.1ex] 
 \vspace{0.1cm}
 Horizontal 2 & $1.2\pm0.4$ & $0.20\pm0.04$ & $6\pm5$ \\[0.1ex] 
 \vspace{0.1cm}
 Harmonics & $1.4\pm0.03$ & $0.16\pm0.03$ & $8\pm9$ \\ [0.1ex] 
\end{tabular}
    
    \label{tab:TLS}
\end{table}

Given these values, we can assume that the nature of resonant TLSs is the same found on other substrates; and since the residual losses are larger than the TLS ones, we can conclude that the resonant TLS are not the main loss mechanism for the resonators.

\subsection{Temperature dependence of $Q_i$ and $f$}
Finally, we study the temperature dependence of the internal quality factor and the resonator frequency. Here, the former provides information on the  quasiparticle density, $n_{qp}$, the latter has two contribution, non-resonant TLSs  and quasiparticles. 

$Q_i$ should decrease with an increase of temperature because $n_{qp}$ increases, leading to a larger resistive loss. Using the loss model from \cite{barends2011minimizing}, $Q_i$ depends linearly on the density $n_{qp}$:
\begin{equation}
  \frac{1}{Q_i}=  \frac{\alpha}{\pi}\sqrt{\frac{2\Delta}{hf_r}}\frac{n_{qp}(T)}{D(E_F)\Delta}+\frac{1}{Q_{B}}, 
  \label{eq:Q_quasiparticles}
\end{equation}
where $\alpha$ is the ratio between kinetic and geometric inductance in the resonator for $k_BT\ll hf_r$, $\Delta$ is the aluminium superconducting gap, $D(E_F)$ is the density of states at the Fermi level, and $Q_{B}$ is the quality factor due to all other losses than quasi particles.
\begin{figure}
\centering
\includegraphics[width=0.47\textwidth]{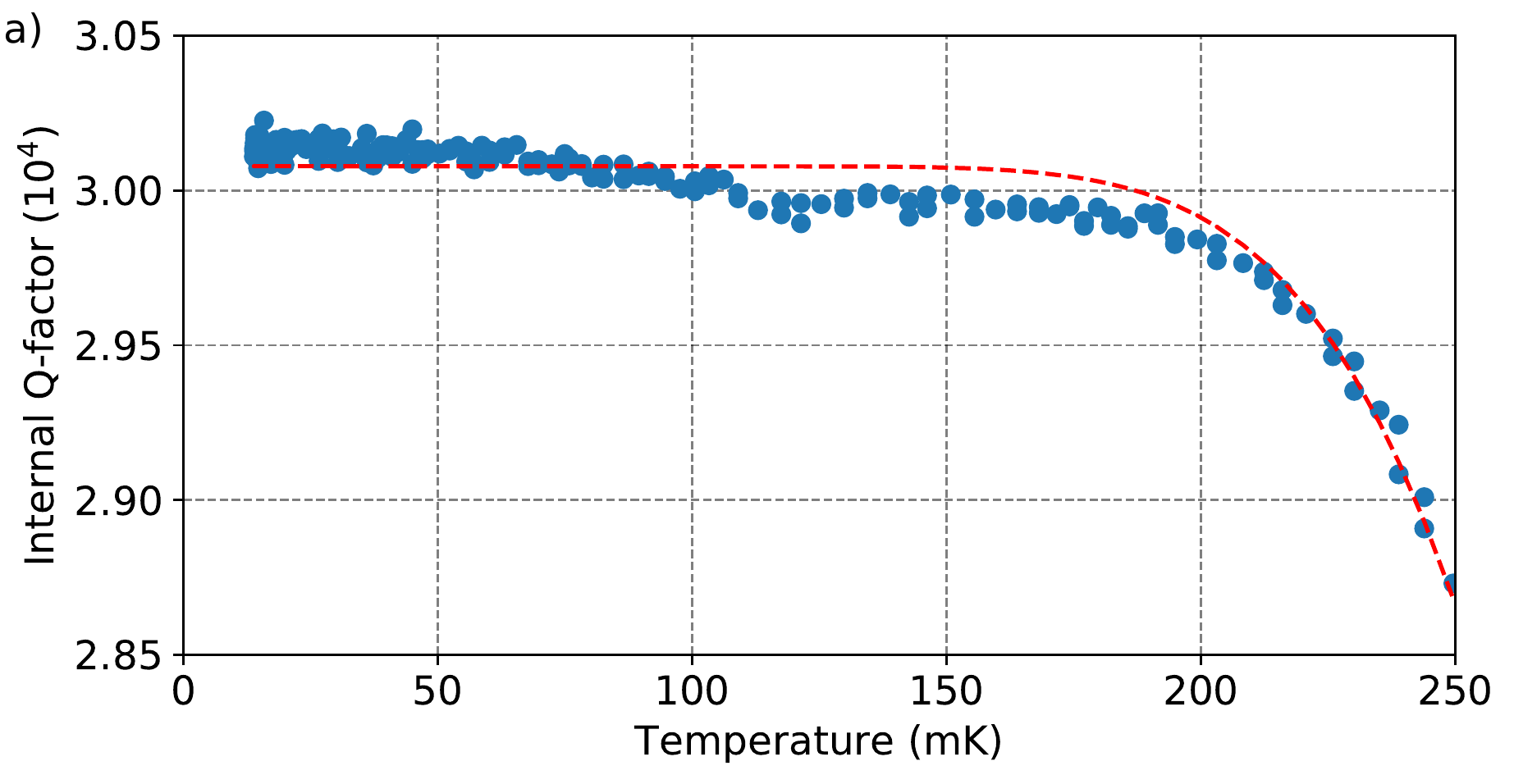}
\includegraphics[width=0.47\textwidth]{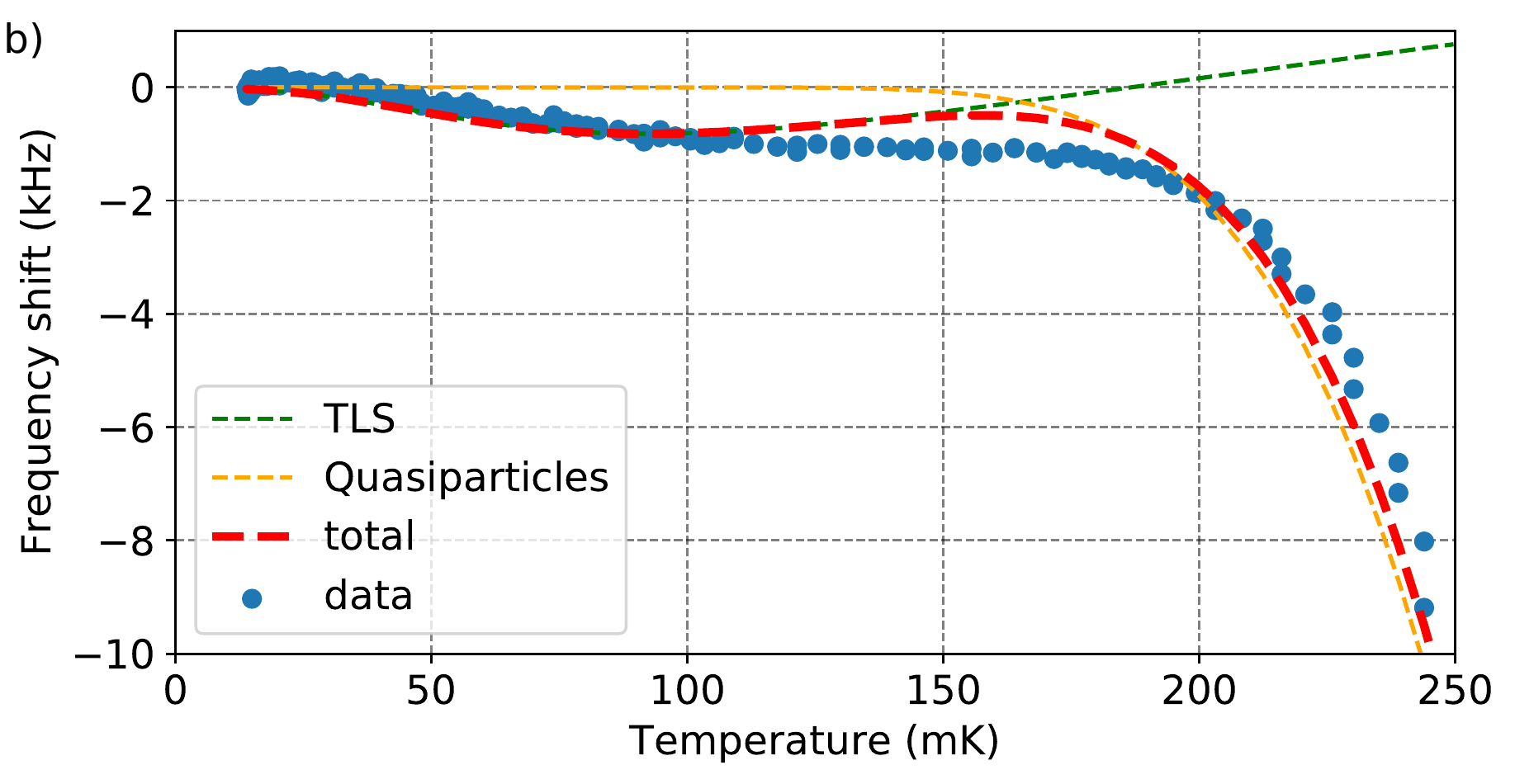}
\caption{Temperature dependence of $Q_i$ and resonance frequency of a resonator on sample Vertical 1: the increase in equilibrium quasiparticles produces a larger resistive loss rate (lower the internal Q) and an additional contribution to the total inductance (lowering the frequency). \textbf{a)} The measured $Q_i$ values are fitted to an equilibrium quasiparticles distribution (red dashed line).\textbf{b)} Frequency shift of $f_r$ as a function of temperature; for $T<100$\,mK the main contribution to the frequency shift is due to non resonant TLS. The  calculation of the TLS frequency shif (formula \ref{eq:TLS_frequency}) is shown in the plot (green dashed lines). For $T>100$\,mK the quasiparticles increase the kinetic inductance and become the main contributor to the frequency shift.}
\label{fig:temp_dependance}
\end{figure}

Figure \ref{fig:temp_dependance}a shows the fit of the measured $Q_i$ for a resonator of sample vertical 1 to formula \ref{eq:Q_quasiparticles}. From the fit we can extract the kinetic inductance ratio $\alpha=1.35\cdot10^{-3}$ and the contributions of other losses $Q_{B}=3.0\cdot 10^{4}$ that dominates the dissipation at lower temperature. Since this value is very close to the numerical evaluation of the mechanical Q factor, we can assume that the quasi particles are not the main loss factor for the resonators at low temperature.

The frequency $f_r$ of the resonators shifts because of two contributions: the first is due to the asymmetrically saturated spectral density that the TLSs give rise to. This effect is described by the TLS loss model \cite{pappas2011two}: 
\begin{equation}
    \Delta f= \frac{1}{Q_{TLS}^0}\frac{1}{\pi}\left[\mbox{Re}\left\{\Psi\left(\frac{1}{2}+\frac{hf_r}{2\pi ik_BT}\right)\right\}-\mbox{ln}\frac{hf_r}{2\pi k_BT}\right],
    \label{eq:TLS_frequency}
\end{equation}
where $\Psi$ is the digamma function. We can plot this function using the $Q_{TLS}^0$ extracted from the previous fit; the results are shown in figure \ref{fig:temp_dependance}b. The second contribution is due to the equilibrium quasiparticle density that increases the kinetic inductance becoming the main factor for the frequency shift:
\begin{equation}
    \Delta f=-\frac{1}{2} f_r\frac{\Delta L}{L}=-\frac{1}{2}\alpha f_r\frac{\Delta L_k}{L_k}.
    \label{eq:kinetic_freq}
\end{equation}

In figure \ref{fig:temp_dependance}b, we plot Eq.\ref{eq:kinetic_freq} using the kinetic inductance ratio obtained from the fit of Eq. \ref{eq:Q_quasiparticles}. The result agrees well with the data for temperature above 100\,mK. The sum of the two contributions is plotted with a red dashed line.
 
\section{Discussion and conclusion}

According to our results, the internal quality factor of CPW resonators on GaAs is mainly limited by the conversion of the photons stored in the resonator to mechanical waves (acoustic phonons) that dissipate energy into the substrate. We find that $Q_i$ increases as a function of frequency, as we numerically predicted. Although a direct measurement of the SAW and BAW generation was not performed, the match between the simulations and the measurements confirms our conclusion. in addition, we exclude the other most common loss sources by measuring the loss as a function of temperature and photon number in the resonator.

We believe that this work can help improving designs and performance of superconducting devices built on piezoelectric substrates. We want to stress that, in the case of superconducting electronics, also in the single photon regime, an oscillating electric field generates mechanical waves due to the electromechanical coupling. Therefore, the lifetime of the excitations in these circuits will be limited by this conversion.

One possible solution to limit this conversion is to design and fabricate phononic mode structures on the substrate to engineer phononic bandgaps. Alternatively, one can minimize the microwave circuitry fabricated on piezoelectric substrates to what is strictly needed in the experiment, either by using two chips in a flip chip geometry \cite{satzinger2018quantum, chu2018creation} or by using piezoelectric thin films which can be selectively etched in designed regions \cite{Chu2017}.


\section{Acknowledgements}
We wish to express our gratitude to Lars J\"{o}nsson for making the sample holder. We gratefully acknowledge the financial support from the Swedish Research council and the Knut and Alice Wallenberg Foundation.

\section{References}

\bibliographystyle{unsrt}
\bibliography{bibliography}

\begin{thebibliography}{10}

\bibitem{megrant2012planar}
A~Megrant, C~Neill, R~Barends, B~Chiaro, Y~Chen, L~Feigl, J~Kelly, E~Lucero,
  M~Mariantoni, P~JJ O’Malley, et~al.
\newblock Planar superconducting resonators with internal quality factors above
  one million.
\newblock {\em Applied Physics Letters}, 100:113510, 2012.

\bibitem{Day2003}
P~K Day, H~G LeDuc, B~A Mazin, A~Vayonakis, and J~Zmuidzinas.
\newblock A broadband superconducting detector suitable for use in large
  arrays.
\newblock {\em Nature}, 425:817, 2003.

\bibitem{Wallraff2004}
A~Wallraff, D~I Schuster, A~Blais, L~Frunzio, R~S Huang, J~Majer, S~Kumar, S~M
  Girvin, and R~J Schoelkopf.
\newblock Strong coupling of a single photon to a superconducting qubit using
  circuit quantum electrodynamics.
\newblock {\em Nature}, 431:162--167, 2004.

\bibitem{goppl2008coplanar}
M~G{\"o}ppl, A~Fragner, M~Baur, R~Bianchetti, S~Filipp, JM~Fink, PJ~Leek,
  G~Puebla, L~Steffen, and Andreas Wallraff.
\newblock Coplanar waveguide resonators for circuit quantum electrodynamics.
\newblock {\em Journal of Applied Physics}, 104:113904, 2008.

\bibitem{wang2009improving}
H~Wang, M~Hofheinz, J~Wenner, M~Ansmann, RC~Bialczak, M~Lenander, Erik Lucero,
  M~Neeley, AD~O’Connell, D~Sank, et~al.
\newblock Improving the coherence time of superconducting coplanar resonators.
\newblock {\em Applied Physics Letters}, 95:233508, 2009.

\bibitem{o2008microwave}
A~D O’Connell, M~Ansmann, R~C Bialczak, M~Hofheinz, N~Katz, E~Lucero,
  C~McKenney, M~Neeley, H~Wang, E~M Weig, et~al.
\newblock Microwave dielectric loss at single photon energies and millikelvin
  temperatures.
\newblock {\em Applied Physics Letters}, 92:112903, 2008.

\bibitem{sage2011study}
J~M Sage, V~Bolkhovsky, W~D Oliver, B~Turek, and P~B Welander.
\newblock Study of loss in superconducting coplanar waveguide resonators.
\newblock {\em Journal of Applied Physics}, 109:063915, 2011.

\bibitem{wisbey2010effect}
D~S Wisbey, J~Gao, M~R Vissers, F~CS da~Silva, J~S Kline, L~Vale, and D~P
  Pappas.
\newblock Effect of metal/substrate interfaces on radio-frequency loss in
  superconducting coplanar waveguides.
\newblock {\em Journal of Applied Physics}, 108:093918, 2010.

\bibitem{sandberg2012etch}
M~Sandberg, M~R Vissers, J~S Kline, M~Weides, J~Gao, David~S W, and D~P Pappas.
\newblock Etch induced microwave losses in titanium nitride superconducting
  resonators.
\newblock {\em Applied Physics Letters}, 100:262605, 2012.

\bibitem{wenner2011wirebond}
J~Wenner, M~Neeley, R~C Bialczak, M~Lenander, E~Lucero, A~D O’Connell,
  D~Sank, H~Wang, M~Weides, A~N Cleland, et~al.
\newblock Wirebond crosstalk and cavity modes in large chip mounts for
  superconducting qubits.
\newblock {\em Superconductor Science and Technology}, 24:065001, 2011.

\bibitem{barends2011minimizing}
R~Barends, J~Wenner, M~Lenander, Y~Chen, R~C Bialczak, J~Kelly, E~Lucero,
  P~O’Malley, M~Mariantoni, D~Sank, et~al.
\newblock Minimizing quasiparticle generation from stray infrared light in
  superconducting quantum circuits.
\newblock {\em Applied Physics Letters}, 99:113507, 2011.

\bibitem{dunsworth2017characterization}
A~Dunsworth, A~Megrant, C~Quintana, Zijun Chen, R~Barends, B~Burkett, B~Foxen,
  Yu~Chen, B~Chiaro, A~Fowler, et~al.
\newblock Characterization and reduction of capacitive loss induced by
  sub-micron josephson junction fabrication in superconducting qubits.
\newblock {\em Applied Physics Letters}, 111:022601, 2017.

\bibitem{PhysRevX.9.021056}
L~R Sletten, B~A Moores, J~J Viennot, and K~W Lehnert.
\newblock Resolving phonon fock states in a multimode cavity with a double-slit
  qubit.
\newblock {\em Physical Review X}, 9:021056, 2019.

\bibitem{satzinger2018quantum}
K~J Satzinger, Y~P Zhong, H-S Chang, G~A Peairs, A~Bienfait, M~Chou, A~Y
  Cleland, C~R Conner, {\'E}~Dumur, J~Grebel, et~al.
\newblock Quantum control of surface acoustic-wave phonons.
\newblock {\em Nature}, 563:661, 2018.

\bibitem{andersson2019non}
G~Andersson, B~Suri, L~Guo, T~Aref, and P~Delsing.
\newblock Non-exponential decay of a giant artificial atom.
\newblock {\em Nature Physics}, pages 1--5, 2019.

\bibitem{chu2018creation}
Y~Chu, P~Kharel, T~Yoon, L~Frunzio, P~T Rakich, and R~J Schoelkopf.
\newblock Creation and control of multi-phonon fock states in a bulk
  acoustic-wave resonator.
\newblock {\em Nature}, 563:666, 2018.

\bibitem{Toida2013}
H~Toida, T~Nakajima, and S~Komiyama.
\newblock Vacuum {R}abi splitting in a semiconductor circuit {QED} system.
\newblock {\em Phys. Rev. Lett.}, 110, February 2013.

\bibitem{Frey2012}
T~Frey, P~J Leek, M~Beck, A~Blais, T~Ihn, K~Ensslin, and A~Wallraff.
\newblock Dipole coupling of a double quantum dot to a microwave resonator.
\newblock {\em Phys. Rev. Lett.}, 108, 2012.

\bibitem{scarlino2019coherent}
P~Scarlino, D~J Van~Woerkom, U~C Mendes, J~V Koski, A~J Landig, C~K Andersen,
  S~Gasparinetti, C~Reichl, W~Wegscheider, K~Ensslin, et~al.
\newblock Coherent microwave-photon-mediated coupling between a semiconductor
  and a superconducting qubit.
\newblock {\em Nature communications}, 10:3011, 2019.

\bibitem{Burnett_2016}
J~Burnett, L~Faoro, and T~Lindstr\"om.
\newblock Analysis of high quality superconducting resonators: consequences for
  {TLS} properties in amorphous oxides.
\newblock {\em Superconductor Science and Technology}, 29:044008, 2016.

\bibitem{probst2015efficient}
S~Probst, FB~Song, PA~Bushev, AV~Ustinov, and M~Weides.
\newblock Efficient and robust analysis of complex scattering data under noise
  in microwave resonators.
\newblock {\em Review of Scientific Instruments}, 86:024706, 2015.

\bibitem{knuuttila2005laser}
J~Knuuttila.
\newblock {\em Laser-interferometric analysis of surface acoustic wave
  resonators}.
\newblock PhD thesis, Helsinki University of Technology, 2005.

\bibitem{gustafsson2012local}
M~V Gustafsson, P~V Santos, G~Johansson, and P~Delsing.
\newblock Local probing of propagating acoustic waves in a gigahertz echo
  chamber.
\newblock {\em Nature Physics}, 8:338, 2012.

\bibitem{doi:10.1177/1045389X18803461}
C~Maruccio, M~Scigliuzzo, S~Rizzato, P~Scarlino, G~Quaranta, M~S Chiriaco, A~G
  Monteduro, and G~Maruccio.
\newblock Frequency and time domain analysis of surface acoustic wave
  propagation on a piezoelectric gallium arsenide substrate: A computational
  insight.
\newblock {\em Journal of Intelligent Material Systems and Structures}, 30:801,
  2019.

\bibitem{1967IBMJ...11..215C}
R~{Courant}, K~{Friedrichs}, and H~{Lewy}.
\newblock {On the Partial Difference Equations of Mathematical Physics}.
\newblock {\em IBM Journal of Research and Development}, 11:215--234, 1967.

\bibitem{Pozar2011}
D~M Pozar.
\newblock {\em Microwave Engineering}.
\newblock John Wiley \& Sons, Inc., 4th ed. edition, 2011.

\bibitem{pappas2011two}
David~P Pappas, Michael~R Vissers, David~S Wisbey, Jeffrey~S Kline, and
  Jiansong Gao.
\newblock Two level system loss in superconducting microwave resonators.
\newblock {\em IEEE Transactions on Applied Superconductivity}, 21:871--874,
  2011.

\bibitem{Burnett_2018}
J~Burnett, A~Bengtsson, D~Niepce, and J~Bylander.
\newblock Noise and loss of superconducting aluminium resonators at single
  photon energies.
\newblock {\em Journal of Physics: Conference Series}, 969:012131, 2018.

\bibitem{PhysRevApplied.11.044014}
D~Niepce, J~Burnett, and J~Bylander.
\newblock High kinetic inductance $\mathrm{Nb}\mathrm{N}$ nanowire
  superinductors.
\newblock {\em Physical Review Applied}, 11:044014, 2019.

\bibitem{Chu2017}
Y~Chu, P~Kharel, W~H Renninger, L~D Burkhart, L~Frunzio, P~T Rakich, and R~J
  Schoelkopf.
\newblock Quantum acoustics with superconducting qubits.
\newblock {\em Science}, 358, 2017.

\end{thebibliography}

\end{document}